\documentclass[11pt]{article}
\usepackage{cite}

\def\crampest{\medmuskip = 1mu plus 1mu minus 1mu}
\def\uncramp{\medmuskip = 4mu plus 2mu minus 4mu}
\def\ben{\begin{equation}}
\def\een{\end{equation}}

  \let\n=\nu

\let\C=\Chi 
\let\la=\label
\newcommand{\eq}[1]{(\ref{#1})}

\def\nn{\nonumber} \def\bd{\begin{document}} \def\ed{\end{document}}
\def\ds{\documentstyle} \let\fr=\frac \let\bl=\bigl \let\br=\bigr
\let\Br=\Bigr \let\Bl=\Bigl
\let\bm=\bibitem
\let\na=\nabla
\let\pa=\partial \let\ov=\overline
\newcommand{\be}{\begin{equation}}
\newcommand{\ee}{\end{equation}}
\def\ba{\begin{array}}
\def\ea{\end{array}}
\def\ft#1#2{{\textstyle{{\scriptstyle #1}\over {\scriptstyle #2}}}}
\def\fft#1#2{{#1 \over #2}}
\def\del{\partial}
\def\vp{\varphi}
\def\sst#1{{\scriptscriptstyle #1}}
\def\oneone{\rlap 1\mkern4mu{\rm l}}
\def\td{\tilde}
\def\wtd{\widetilde}
\def\ie{\rm i.e.\ }
\def\dalemb#1#2{{\vbox{\hrule height .#2pt
        \hbox{\vrule width.#2pt height#1pt \kern#1pt
                \vrule width.#2pt}
        \hrule height.#2pt}}}
\def\square{\mathord{\dalemb{6.8}{7}\hbox{\hskip1pt}}}
\newcommand{\ho}[1]{$\, ^{#1}$}
\newcommand{\hoch}[1]{$\, ^{#1}$}
\newcommand{\bea}{\begin{eqnarray}}
\newcommand{\eea}{\end{eqnarray}}
\newcommand{\ra}{\rightarrow}
\newcommand{\lra}{\longrightarrow}
\newcommand{\Lra}{\Leftrightarrow}
\newcommand{\ap}{\alpha^\prime}
\newcommand{\bp}{\tilde \beta^\prime}
\newcommand{\tr}{{\rm tr} }
\newcommand{\Tr}{{\rm Tr} }
\def\0{{\sst{(0)}}}
\def\1{{\sst{(1)}}}
\def\2{{\sst{(2)}}}
\def\3{{\sst{(3)}}}
\def\4{{\sst{(4)}}}
\def\5{{\sst{(5)}}}
\def\6{{\sst{(6)}}}
\def\7{{\sst{(7)}}}
\def\8{{\sst{(8)}}}
\def\n{{\sst{(n)}}}
\def\cA{{{\cal A}}}
\def\cF{{{\cal F}}}
\def\tV{\widetilde V}
\def\tW{\widetilde W}
\def\tH{\widetilde H}
\def\tE{\widetilde E}
\def\tF{\widetilde F}
\def\tA{\widetilde A}
\def\im{{{\rm i}}}
\def\tY{{{\wtd Y}}}
\def\ep{{\epsilon}}
\def\vep{{\varepsilon}}
\def\R{\rlap{\rm I}\mkern3mu{\rm R}}
\def\bD{{{\bar D}}}

\def\R{\rlap{\rm I}\mkern3mu{\rm R}}
\def\bD{{{\bar D}}}
\def\R{{{\Bbb R}}}
\def\C{{{\Bbb C}}}
\def\H{{{\Bbb H}}}
\def\CP{{{\Bbb C}{\Bbb P}}}
\def\RP{{{\Bbb R}{\Bbb P}}}
\def\Z{{{\Bbb Z}}}
\def\bA{{{\Bbb A}}}
\def\bB{{{\Bbb B}}}
\def\bC{{{\Bbb C}}}
\def\bR{{{\Bbb R}}}
\def\bD{{{\Bbb D}}}
\def\bE{{{\Bbb E}}}
\def\bZ{{{\Bbb Z}}}
\def\Re{{{\frak{Re}}}}
\def\Im{{{\frak{Im}}}}
\def\cosec{{\,\hbox{cosec}\,}}
\def\Gm{{\Gamma_{\!\! -}}}
\def\Gp{{\Gamma_{\!\! +}}}
\def\stan{{standard }}
\def\nonstan{{supernumerary }}

\def\cosech{{\hbox{cosech}}}

\def\etcyc{{\hbox{and cyclic}}}
\def\btheta{{\bar\theta}}

\newcommand{\istanbul}{\it Department of Mathematics,  Bo{\u g}azi{\c c}i
University, Bebek, Istanbul 34342, Turkey.}

\newcommand{\tamphys}{\it George P. \& Cynthia W. Mitchell Institute for
Fundamental Physics,\\
Texas A\&M University, College Station, TX 77843--4242, USA}
\newcommand{\umich}{\it Michigan Center for Theoretical Physics,
University of Michigan\\ Ann Arbor, MI 48109--1120, USA}
\newcommand{\upenn}{\it Department of Physics and Astronomy,
University of Pennsylvania, Philadelphia,  PA 19104, USA}
\newcommand{\SISSA}{\it  SISSA-ISAS and INFN, Sezione di Trieste\\
Via Beirut 2-4, I-34013, Trieste, Italy}

\newcommand{\newton}{\it Isaac Newton Institute for Mathematical Sciences,\\
20 Clarkson Road,  University of Cambridge,
Cambridge CB3 0EH, UK}

\newcommand{\ihp}{\it Institut Henri Poincar\'e\\
  11 rue Pierre et Marie Curie, F 75231 Paris Cedex 05}

\newcommand{\damtp}{\it DAMTP, Centre for Mathematical Sciences,
 Cambridge University\\  Wilberforce Road, Cambridge CB3 OWA, UK}
\newcommand{\itp}{\it Institute for Theoretical Physics, University of
California\\ Santa Barbara, CA 93106, USA}

\newcommand{\auth}{R. G\"uven\hoch{\dagger 1},
James T. Liu\hoch{\ddagger2},
C.N. Pope\hoch{\star3} and E. Sezgin\hoch{\star4} }

\begin{document}
\begin{flushright}
\hfill{
MCTP-03-29\ \
MIFP-03-13}\\
\hfill{
\bf hep-th/0306201}
\end{flushright}

\begin{center}

{\large {\bf Fine Tuning and Six-Dimensional Gauged $N=(1,0)$ Supergravity
Vacua}}

\vspace{15pt}

\auth

\vspace{7pt}
{\hoch{\dagger}\istanbul}

\vspace{7pt}
{\hoch{\ddagger}\umich}

\vspace{7pt}
{\hoch{\star}\tamphys}

\underline{ABSTRACT}
\end{center}

   We find a new family of supersymmetric vacuum solutions in the
six-dimensional chiral gauged $N=(1,0)$ supergravity theory.  They are
generically of the form AdS$_3\times S^3$, where the 3-sphere is
squashed homogeneously along its Hopf fibres.  The squashing is freely
adjustable, corresponding to changing the 3-form charge, and the
solution is supersymmetric for all squashings.  In a limit where the
length of the Hopf fibres goes to zero, one recovers, after a
compensating rescaling of the fibre coordinate, a solution that is
locally the same as the well-known (Minkowski)$_4\times S^2$ vacuum of
this theory.  It can now be viewed as a fine tuning of the new more
general family.  The traditional ``Cosmological Constant Problem'' is
replaced in this theory by the problem of why the four-dimensional
(Minkowski)$_4\times S^2$ vacuum should be selected over other members
of the equally supersymmetric AdS$_3\times S^3$ family.  We also
obtain a family of dyonic string solutions in the gauged $N=(1,0)$
theory, whose near-horizon limits approach the AdS$_3$ times squashed
$S^3$ solutions.

{\vfill\leftline{}\vfill
\vskip 10pt \footnoterule
{\footnotesize \hoch{1}
Research supported in part by the Turkish Academy of Sciences (TUBA)
\vskip -12pt} \vskip 14pt
{\footnotesize \hoch{2}
Research supported in part by DOE grant DE-FG02-95ER40899
\vskip -12pt} \vskip 14pt
{\footnotesize \hoch{3}
Research supported in part by DOE grant DE-FG03-95ER40917
\vskip -12pt} \vskip 14pt
{\footnotesize \hoch{4}
Research supported in part by NSF grant PHY-0070964
\vskip -12pt}  \vskip  14pt
}

\pagebreak
\setcounter{page}{1}


\section{Introduction}

   Six-dimensional chiral supergravities with eight real supersymmetries
admitting spontaneous
compactifications to four-dimensional Minkowski spacetimes have
received both early
\cite{salsez,Koh:1984zf,Maeda:1984gq,Maeda:1985es,Halliwell:1986bs}
and more recent \cite{Carroll:2003db,Navarro:2003vw,Aghababaie:2003wz}
attention as possible scenarios for achieving a naturally small
cosmological constant.  The starting point for a minimal model of this type
is a six-dimensional chiral $N=(1,0)$ Einstein-Maxwell supergravity
with an exponential potential for the dilaton, which results from the
gauging of a $U(1)$ R-symmetry. In the absence of branes, compactification
to four dimensions proceeds by turning on a monopole configuration on a
two-sphere, resulting in a geometry of the form (Minkowski)$_4\times
S^2$ \cite{salsez}.

    In general, achieving a small four-dimensional cosmological
constant requires a fine tuning to balance the bulk vacuum energy with
the monopole flux.\footnote{For example see \cite{rasast}, where
six-dimensional Einstein-Maxwell gravity with an arbitrary
cosmological constant is compactified to $M_4\times S^2$, where $M_4$
is Minkowski, de Sitter or anti-de Sitter spacetime, depending on the
fine tuning of the $S^2$ monopole charge relative to the 6D
cosmological constant.}  However, it has been argued
\cite{Koh:1984zf,Maeda:1984gq,Maeda:1985es} that a form of self-tuning
occurs naturally in the gauged chiral $N=(1,0)$ supergravity of
\cite{salsez}.  The bosonic sector of this theory includes a 2-form
potential and a dilaton, in addition to the metric and a Maxwell
field.  Self-tuning in this theory occurs because the cosmological
constant is replaced by an exponential potential for the dilaton
field.  In this case, the only possibility for a maximally-symmetric
solution turns out to be (Minkowski)$_4\times S^2$ \cite{salsez}, thus
selecting out a Minkowski vacuum without the need for any apparent
fine tuning.  The vacuum yields $N=1$ supergravity in four dimensions
with chiral matter \cite{salsez}.

    In the framework of \cite{Carroll:2003db,Navarro:2003vw,Aghababaie:2003wz},
this basic scenario is extended by introducing 3-branes into the
(Minkowski)$_4\times S^2$ background, giving rise to (American)
``football-shaped'' extra dimensions.  Being co-dimension two objects,
the only effect of the 3-branes on the background is to introduce deficit
angles on the $S^2$.  In particular, the four-dimensional cosmological
constant remains zero (or at least small), provided it is initially
vanishing in the absence of branes.  It is remarkable that this is true
regardless of whether supersymmetry is preserved or broken on the 3-branes
themselves.  Thus we could be living on a braneworld without supersymmetry
and yet still experience a naturally small cosmological constant.
Essentially what happens in these models is that the cosmological constant
problem is transformed into that of finding a naturally stable bulk solution
admitting a flat Minkowski background.

   Since the (Minkowski)$_4\times S^2$ background is the starting
point for the braneworld models with football-shaped extra dimensions,
it is important to address the argued uniqueness of this background in
the context of the $N=(1,0)$ gauged supergravity theory.  In this
paper, we show that there is in fact a more general class of $N=1$
supersymmetric vacuum solutions.  We find a family of solutions of the
form AdS$_3\times S^3$, where $S^3$ is the 3-sphere with a
homogeneously squashed metric.  There is a non-trivial parameter in
the family of solutions, which allows the length $L$ of $U(1)$ fibres
in the description of $S^3$ as the Hopf bundle over $S^2$ to be
adjusted freely, relative to the size of the $S^2$ base.  In the
singular limit where the length $L$ goes to zero one can recover,
after making an appropriate rescaling of the fibre coordinate, a
solution that is locally the same as the previous (Minkowski)$_4
\times S^2$ solution found in \cite{salsez}.  The new feature in the
more general family of AdS$_3\times S^3$ solutions is that in addition
to the 2-form ``monopole flux,'' there are also electric and magnetic
charges carried by the 3-form field.

   In the light of our new solutions, all of which are supersymmetric,
it can be argued that the ostensible absence of fine tuning in the
(Minkowski)$_4\times S^2$ solution of \cite{salsez} is somewhat
deceptive.  In fact there {\it is} a fine tuning, in that it is the
special case of our new supersymmetric vacua in which the electric and
magnetic 3-form charges are set to zero.  The familiar notion of
fine-tuning a maximally-symmetric vacuum solution to have vanishing
cosmological constant is now replaced by a rather less familiar type
of fine tuning.  If the 3-form charges are allowed to become
non-zero not only does one lose the feature of a vanishing cosmological
constant, but one also loses a dimension.  Namely, one of the three
spatial dimensions of the previous (Minkowski)$_4$ vacuum acquires a
non-trivial ``twist'' and becomes the Hopf fibre coordinate of a
3-sphere, whose $S^2$ base formed the original internal space in the
Minkowski vacuum.

   The celebrated ``cosmological constant problem'' is thus replaced,
in this theory, by the ``Hopf fibration problem.''

   After constructing the family of AdS$_3$ times squashed $S^3$
solutions we then show that they can be viewed as the near-horizon
limits of a family of dyonic strings in the six-dimensional
gauged $N=(1,0)$ supergravity.  These solutions can be viewed as
generalisations of the usual dyonic strings of the ungauged theory
\cite{dufekhra}.  As in those examples, the dyonic strings that we
find here preserve one quarter of the original six-dimensional
supersymmetry.  As one reaches the AdS$_3$ times squashed $S^3$
near-horizon limit, the supersymmetry fraction increases from one
quarter to one half in the standard way.

\section{The Six-Dimensional $N=(1,0)$ Gauged Supergravity}

    A large class of six-dimensional $N=(1,0)$ gauged supergravities
has been constructed
\cite{Nishino:1984gk,Nishino:dc,Nishino:1997ff,Ferrara:1997gh,Riccioni:2001bg}.
In this paper, we shall be focusing on
the simplest example, for which the field content comprises a
graviton multiplet with bosonic fields $(g_{MN},B^+_{MN})$ and
chiral (complex) gravitino superpartner $\psi_M$; a tensor multiplet
with bosonic field $B^-_{MN}$ and chiral spin$-\ft12$ superpartner $\chi$;
and a vector multiplet with bosonic field $A_M$ and chiral superpartner
$\lambda$ \cite{salsez}.

    The bosonic sector of the six-dimensional $N=(1,0)$ gauged supergravity
is described by the Lagrangian
\be
{\cal L} = R\, {*\oneone} - \ft14 {*d\phi}\wedge d\phi  - \ft12 e^\phi\,
{*H_\3}\wedge H_\3 - \ft12 e^{\ft12\phi}\, {*F_\2}\wedge F_\2
 - 8g^2 \, e^{-\ft12\phi}\, {*\oneone}\,,
\ee
where $F_\2= dA_\1$, $H_\3 = dB_\2 + \ft12 F_\2\wedge A_\1$, and $g$ is
the gauge-coupling constant.  This leads to the bosonic equations of motion
\bea
R_{MN} &=& \ft14 \del_M\phi\, \del_N\phi + \ft12 e^{\ft12\phi}\,
(F^2_{MN} - \ft18 F^2\,g_{MN}) + \ft14 e^\phi\,
(H^2_{MN} - \ft16 H^2\, g_{MN})\nn\\
&& + 2g^2\,  e^{-\ft12\phi}\ g_{MN}\,,\nn\\
\square\, \phi &=& \ft14 e^{\ft12\phi}\, F^2 + \ft16 e^\phi\, H^2 -
    8g^2\, e^{-\ft12\phi}\,,\nn\\
d(e^{\ft12\phi}\,{*F_\2}) &=& e^\phi\, {*H_\3}\wedge F_\2\,,\label{eoms}\\
d(e^\phi\, {*H_\3}) &=& 0\,.\nn
\eea

    The transformations rules for the fermionic fields are given by%
\footnote{Note that, compared to \cite{salsez}, we have chosen units in
which the gravitational coupling constant $\kappa=1/2$, and have additionally
rescaled the supersymmetry parameter $\epsilon$.}
%
\bea
\delta\psi_M &\equiv&\tilde D_M\ep =
[D_M + \ft1{48} e^{\ft12\phi}\, H^+_{NPQ}\,\Gamma^{NPQ}\,\Gamma_M]\ep\,,\nn\\
\delta\chi &\equiv&-\ft14\Delta_\phi\ep=
-\ft14[\Gamma^M\del_M\phi - \ft16 e^{\ft12\phi}\, H^-_{MNP}\,
\Gamma^{MNP}]\ep\,,\label{susytrans}\\
\delta \lambda &\equiv& \ft1{4\sqrt{2}}\Delta_F\ep=
\ft1{4\sqrt{2}}[e^{\ft14\phi}\, F_{MN}\, \Gamma^{MN} - 8\im\,
g\, e^{-\ft14\phi}]\ep\,,\nn
\eea
where $D_M$ is the gauge-covariant derivative, $D_M\ep \equiv (\nabla_M -
\im\, g\, A_M)\ep$.
Note that the $\pm$ superscripts appearing on the 3-form $H_{MNP}$ in
these expressions are redundant, since the chirality of $\ep$ already
implies projections onto the self-dual or anti-self-dual parts,
but we include them for convenience, to emphasise which projection
occurs in which transformation rule.

\section{Vacuum Solution}
\label{sec:vacsol}

    In this section, we show that the $N=(1,0)$ gauged supergravity
theory admits a one-parameter family of supersymmetric solutions that
generically are of the form AdS$_3\times S^3$.  The non-trivial
parameter in the solution characterises the degree of ``squashing'' of
the internal 3-sphere, which is viewed as the Hopf bundle over $S^2$.
For a particular limiting value of the parameter, in which the length
of the $U(1)$ Hopf fibres tends to zero, the solution locally
approaches, after an appropriate rescaling of the shrinking Hopf
fibres, the (Minkowski)$_4\times S^2$ solution found long ago in
\cite{salsez}.

    The construction of our family of AdS$_3\times S^3$ solutions
proceeds straightforwardly.  We make the following ansatz for the
metric and other bosonic fields:
\bea
ds_6^2 &=& ds_3^2 + a^2\, (\sigma_1^2 + \sigma_2^2) + b^2\, \sigma_3^2\,,\nn\\
F_\2 &=& k\, \sigma_1\wedge \sigma_2\,,\label{vacans}\\
H_\3 &=& P \, \sigma_1\wedge\sigma_2\wedge \sigma_3 + \fft{P}{a^2\,
b}\, \varepsilon_\3\,.\nn
\eea
Here, $a$, $b$, $k $ and $P$ are constants, and the $\sigma_i$ are
left-invariant 1-forms on the 3-sphere, satisfying the exterior algebra
\be
d\sigma_i = -\ft12 \ep_{ijk}\, \sigma_j\wedge \sigma_k\,.
\ee
They can be represented, in terms of Euler angles $(\theta,\varphi,\psi)$,
by
\be
\sigma_1 + \im\, \sigma_2 = e^{-\im\, \psi}\, (d\theta + \im\, \sin\theta\,
d\varphi)\,,\qquad \sigma_3 = d\psi +  \cos\theta\, d\varphi\,.
\ee
The 3-form $\varepsilon_\3$ appearing in the ansatz for $H_\3$ in
(\ref{vacans}) denotes the volume form in the metric $ds_3^2$.  Note
that $H_\3$ in (\ref{vacans}) is explicitly constructed to be self-dual.
Locally, we can choose the potential for $F_\2$ to be given by $A_\1=-k\,
\sigma_3$.

    An elementary calculation shows that in the natural vielbein
basis $e^1=a\, \sigma_1$, $e^2=a\, \sigma_2$, $e^3=b\, \sigma_3$, the
torsion-free spin connection for the $S^3$ factor in the metric $ds_6^2$ in
(\ref{vacans}) is given by
\be
\omega_{23} = - \fft{b}{2a^2}\,  e^1\,,\qquad
\omega_{31} = -\fft{b}{2a^2}\, e^2\,,\qquad
\omega_{12} = \Big(\fft{b}{2a^2} - \fft1{b}\Big)\, e^3\,,
\ee
and the non-vanishing vielbein components of the Ricci tensor are given by
\be
R_{11}=R_{22} = \fft1{a^2} - \fft{b^2}{2a^4}\,,\qquad
R_{33} = \fft{b^2}{2a^4}\,.
\ee

   The equation of motion for $H_\3$ in (\ref{eoms}) implies, in
view of the ansatz (\ref{vacans}), that $\phi$ is a constant, which
without loss of generality we can take to be zero.  The equations of
motion then lead to the following relations:
\be
k^2=16 g^2\, a^4\,,\qquad b^2=P\,,\qquad a^2 = b^2 + \ft12
k^2\,,\label{ads3sol1}
\ee
together with
\be
R_{\mu\nu} = -\fft{b^2}{2a^4}\, g_{\mu\nu}\,.\label{adsric}
\ee
Thus we can view the 3-form charge $P$ as a free parameter, with the
remaining parameters in the ansatz (\ref{vacans}) determined by
\be
b^2 = P\,,\qquad a^2 = \fft{k}{4g}\,,\qquad k\, (1-2g\,k)
  = 4g\, P
\,.\label{ads3sol3}
\ee
We have, without loss of generality, made a specific sign choice when
taking the square root of the first equation in (\ref{ads3sol1}).
When $P$ is non-vanishing, the solution is of the form AdS$_3\times
S^3$, and since the ratio $b/a$ is not equal to 1, the metric on the
$S^3$ is squashed along the Hopf fibres.  The equation determining $a^2$
has two branches, with
\be
a^2 = \fft1{16 g^2}\, \Big(1 \pm \sqrt{1-32P\, g^2}\Big)\,.\label{asol}
\label{reln1}
\ee

   If we choose the $+$ sign in (\ref{asol}), then $a$ is
non-vanishing in the limit $P\longrightarrow0$, while $b$ tends to
zero.  If the Euler angle $\psi$ is rescaled to $\psi=b^{-1}\, z$
before taking the limit, then the metric takes the form
\be
ds_6^2 = ds_3^2 + a^2\, (d\theta^2 + \sin^2\theta\, d\varphi^2) +
               (dz + b\, \cos\theta\, d\varphi)^2\,.
\ee
If $b$ goes to zero, we see from (\ref{adsric}) that $ds_3^2$ becomes
Ricci flat.  This leads to an ``untwisting'' of the Hopf fibre
coordinate $z$, leaving in the limit a metric that is the direct sum
of the metric on $S^2$ and a Ricci-flat 4-metric, which could be taken
to be (Minkowski)$_4$.  Thus one recovers the (Minkowski)$_4\times
S^2$ solution of \cite{salsez} as a special limiting case of our
family of AdS$_3$ times squashed $S^3$ solutions, as the 3-form charge
$P$ is sent to zero.\footnote{Since the $U(1)$ fibre coordinate $\psi$
on $S^3$ has period $4\pi$, it follows that the period of the rescaled
coordinate $z$ will tend to zero as $b=\sqrt{P}$ goes to zero. Once the limit
$b=0$ is reached, one can ``unwrap'' the collapsed circle and allow
$z$ to range over the entire real line, thus giving a
(Minkowski)$_4\times S^2$ topology.  An alternative viewpoint would be
to start from the ``fine-tuned'' (Minkowski)$_4\times S^2$ solution at
$P=0$, using the coordinate $z$ ranging over the entire real line.
When $P$ is then allowed to become non-zero, one would encounter
conical singularities in the solution, since the coordinate $z$ would
be covering the $U(1)$ fibre in $S^3$ infinitely many times (and the
3-form charge $\int H_\3 = 4\pi\, \sqrt{P}\, (\Delta z)$ would be
infinite owing to the infinite period $(\Delta z)$ for $z$).  These
solutions can then be made regular, giving
AdS$_3\times S^3$, by restricting $z$ to have period $(\Delta z) =
4\pi\, \sqrt{P}$.}  The (Minkowski)$_4 \times S^2$ solution
can therefore be viewed as a fine-tuning in which the 3-form charge
$P$ is set to zero, implying that the strength of the 2-form flux
takes the specific value
\be
k = \fft1{2g}\,.
\ee

   It is a straightforward matter to verify that our solution is
supersymmetric for all values of $P$.  First of all, we see from
(\ref{susytrans}) that we shall have $\delta\, \chi=0$, since $\phi=0$
and $H_\3$ is self-dual.  In fact, the tensor multiplet is completely
decoupled from this solution.  Next, we see that $\delta\, \lambda=0$
implies
\be
\fft{k}{a^2}\, \Gamma_{12}\, \ep = 4\im\, g\, \ep\,,
\ee
which requires, since we chose $k=+4 g\, a^2$, that
\be
\Gamma_{12}\, \ep = + \im\, \ep\,.\label{gam12proj}
\ee
Note that the Dirac matrices $\Gamma_1$, $\Gamma_2$ and $\Gamma_3$ are
given with frame indices
corresponding to the $S^3$.  Condition (\ref{gam12proj}) implies a
halving of the original six-dimensional supersymmetry.  Turning finally
to the gravitino variation, from
$\delta\, \psi_M=0$ with $M$ in the $S^3$ directions, we find for
$\delta\, \psi_1=0$ and $\delta\, \psi_2=0$ that $b^2=P$, which is
consistent with (\ref{ads3sol1}).  From $\delta\, \psi_3=0$ we find
$b^2=a^2\, (1-2g\, k)$, which, since $k=4\, g\, a^2$, is also consistent
with (\ref{ads3sol1}).  Note that we have $\del_i\, \ep=0$, \ie the
Killing spinors are independent of the coordinates of $S^3$.  Finally
from $\delta\, \psi_\mu=0$ in the AdS$_3$ directions we obtain
\be
\nabla_\mu \, \ep = -\fft{\im\, P}{4 a^2\, b}\, \Gamma_3\, \Gamma_\mu\,
\ep = -\ft{\im}{2}\, \sqrt{-\fft{\Lambda}{2}}\, \Gamma_3\, \Gamma_\mu\, \ep\,,
\label{adsks}
\ee
where $\Lambda \equiv -P^2/(2a^4\, b^2) = -b^2/(2a^4)$ is, from
(\ref{adsric}), the cosmological constant of the AdS$_3$
spacetime.  Equation (\ref{adsks}) is nothing but a statement that
$\ep$ must be a Killing spinor in AdS$_3$.  This completes the
demonstration that our AdS$_3\times S^3$ solutions preserve one half
of the original six-dimensional supersymmetry.

\section{Dyonic String Solutions in $N=(1,0)$ Gauged Supergravity}

    Having shown that the $N=(1,0)$ gauged supergravity admits a family
of AdS$_3$ times squashed $S^3$ solutions, it is natural to ask whether
such vacua may be related to string-like objects.  In this section we
show that this is indeed the case.  In particular, we construct dyonic
string solutions preserving $\ft14$ of the original six-dimensional
supersymmetry.  In the near-horizon limit, these strings approach the
AdS$_3$ times squashed $S^3$ solutions of the previous section, whereupon
supersymmetry is partially restored from $\ft14$ to $\ft12$ of the
original supersymmetry.

\subsection{The equations of motion, and Killing-spinor conditions}

    Our starting point is the following ansatz for string solutions:
\bea
ds_6^2 &=& c^2\, dx^\mu\, dx_\mu + a^2\, (\sigma_1^2 + \sigma_2^2) +
       b^2\, \sigma_3^2 + h^2\, dr^2\,,\nn\\
H_\3 &=& P\, \sigma_1\wedge \sigma_2\wedge\sigma_3 + u\, d^2x\wedge dr
\,,\\
F_2 &=& k\, \sigma_1\wedge \sigma_2\,,\nn
\eea
where $a$, $b$, $c$, $h$, $u$ and $\phi$ are now all functions of $r$.
The magnetic charge $P$ and the coefficient $k$ are constants, by
virtue of the Bianchi identities for $H_\3$ and $F_\2$.  While $h$ may be
removed by a coordinate transformation, $dr'=h(r)\,dr$, the string solution
simplifies for a suitable choice of $h$ (as will be apparent below).  For
this reason we will retain $h$ in the following.
When it is necessary to use explicit
numerical vielbein component labels, we use the following basis:
\be
e^{\td 0} = c\, dt\,,\quad e^{\td 1} = c\, dx\,,\quad
e^1= a\, \sigma_1\, \quad e^2=a\, \sigma_2\, \quad
e^3=b\, \sigma_3\,,\quad
e^4=h\, dr\,.
\ee
In this orthonormal frame, the non-vanishing components of the
spin connection are given by
\bea
&&\omega_{23} = -\fft{b}{2a^2}\, e^1\,,\quad
\omega_{31} = -\fft{b}{2 a^2}\, e^2\,,\quad
\omega_{12} = \left(\fft{b}{2a^2} - \fft1{b}\right)\, e^3\,,\nn\\
&&\omega_{14}= \fft{a'}{a\, h}\, e^1\,,\quad
\omega_{24}= \fft{a'}{a\, h}\, e^2\,,\quad
\omega_{34}= \fft{b'}{b\, h}\, e^3\,,\quad
\omega^\mu{}_4= \fft{c'}{c\, h}\, e^\mu\,,
\eea
and the non-vanishing components of the Ricci tensor are given by
\bea
R_{\mu\nu} &=& -\Big[
\fft{{c'}^2}{h^2\, c^2} + \fft{2 a'\, c'}{a\, c\, h^2} +
\fft{b'\, c'}{b\, c\, h^2}  + \fft1{c\, h}\, \Big(\fft{c'}{h}\Big)'
\Big]\, \eta_{\mu\nu}\,,\nn\\
R_{11}&=& R_{22} = -\fft{2 a'\, c'}{a\, c\, h^2} -
\fft{a'\, b'}{a\, b\, h^2} -
\fft{{a'}^2}{a^2\, h^2} - \fft1{a\, h}\, \Big( \fft{a'}{h}\Big)'
  - \fft{b^2}{2a^4} + \fft1{a^2}\,,\nn\\
R_{33}&=& -\fft{2 b'\, c'}{a\, c\, h^2} - \fft{2 a'\, b'}{a\, b\, h^2}
 -  \fft1{b\, h}\, \Big( \fft{b'}{h}\Big)'
  + \fft{b^2}{2a^4}\,,\nn\\
R_{44} &=& -\fft{2}{a\, h}\, \Big(\fft{a'}{h}\Big)' -
\fft{1}{b\, h}\, \Big(\fft{b'}{h}\Big)' -
\fft{2}{c\, h}\, \Big(\fft{c'}{h}\Big)'\,.
\eea

   The field equations for $F_\2$ and $H_\3$ given in (\ref{eoms}) imply
\be
b^2 = P\, e^{\ft12\phi}\,,\qquad u = \fft{h\, c^2\, Q}{a^2\, b}\,
e^{-\phi}\label{bsol}
\ee
respectively, where $Q$ is a constant characterising the electric
charge carried by $H_\3$.  The Einstein and dilaton equations of motion give
\crampest \bea \fft{{c'}^2}{h^2\, c^2} + \fft{2 a'\, c'}{a\, c\,
h^2} + \fft{b'\, c'}{b\, c\, h^2}  + \fft1{c\, h}\,
\Big(\fft{c'}{h}\Big)' \!&=&\! \fft{k^2}{8 a^4}\, e^{\ft12\phi}\,
+ \ft14 e^\phi\, \Big( \fft{u^2}{h^2\, c^4} + \fft{P^2}{a^4\,
b^2}\Big) - 2 g^2\, e^{-\ft12\phi}
\,,\nn\\
\fft{2 a'\, c'}{a\, c\, h^2} + \fft{a'\, b'}{a\, b\, h^2} +
\fft{{a'}^2}{a^2\, h^2} + \fft1{a\, h}\, \Big( \fft{a'}{h}\Big)'
  + \fft{b^2}{2a^4} - \fft1{a^2}
\!&=&\! -\fft{3 k^2}{8 a^4}\, e^{\ft12\phi} -\ft14 e^{\phi}\,
\Big( \fft{u^2}{h^2\, c^4} + \fft{P^2}{a^4\, b^2}\Big) - 2 g^2
 e^{-\ft12\phi} \!,\nn\\
\fft{2 b'\, c'}{a\, c\, h^2} + \fft{2 a'\, b'}{a\, b\, h^2} +
  \fft1{b\, h}\, \Big( \fft{b'}{h}\Big)'
  - \fft{b^2}{2a^4}
\!&=&\! \fft{k^2}{8 a^4}\, e^{\ft12\phi} -\ft14 e^{\phi}\,  \Big(
\fft{u^2}{h^2\, c^4} + \fft{P^2}{a^4\, b^2}\Big) - 2 g^2\,
e^{-\ft12\phi}
\,,\nn\\
\fft{2}{a\, h}\, \Big(\fft{a'}{h}\Big)' +
\fft{1}{b\, h}\, \Big(\fft{b'}{h}\Big)' +
\fft{2}{c\, h}\, \Big(\fft{c'}{h}\Big)' + \fft{{\phi'}^2}{4h^2}
\!&=&\!
 \fft{k^2}{8 a^4}\, e^{\ft12\phi} +
    \ft14 e^\phi\, \Big(
\fft{u^2}{h^2\, c^4} + \fft{P^2}{a^4\, b^2}\Big)
- 2 g^2\, e^{-\ft12\phi}\,,\nn\\
\fft{1}{a^2\, b\, c^2\, h}\, \Big(\fft{a^2\, b\, c^2\,
\phi'}{h}\Big)' \!&=&\! \fft{k^2}{2a^4}\, e^{\ft12\phi} + e^\phi\,
\Big( \fft{P^2}{a^4\, b^2} - \fft{u^2}{h^2\, c^4} \Big)-8g^2\,
e^{-\ft12\phi}\,.
\nonumber\\
\label{eoms2}
\eea
\uncramp

    Rather than trying to solve the rather complicated second-order
Einstein and dilaton equations of motion, we shall look directly for
supersymmetric solutions by studying the conditions for the existence
of Killing spinors.  Having found such configurations, we then verify
that they do indeed satisfy the second-order field equations.
First, the condition $\delta\lambda=0$ implies
\be
\fft{k}{a^2}\, \Gamma_{12}\, \ep = 4\im\, g\, e^{-\ft12\phi}\, \ep\,.
\ee
Without loss of generality, we can require that $\ep$ satisfy the same
projection condition (\ref{gam12proj}) as in section \ref{sec:vacsol},
and hence we have
\be
a^2 = \fft{k}{4g}\, e^{\ft12\phi}\,.\label{asol2}
\ee
Next, from $\delta\chi=0$, we find using (\ref{susytrans}) that
\be
\phi'\, \Gamma_4\, \ep = \fft{h}{a^2\, b}\, e^{\ft12\phi}\, (P- Q\,
e^{-\phi})\, \Gamma_{123}\, \ep\,,
\ee
where $Q$ is given in (\ref{bsol}).
This implies a further halving of supersymmetry by a projection condition
which, without loss of generality, we take to be
\be
\Gamma_{1234}\, \ep = + \ep\,,\label{gam1234proj}
\ee
and hence we have the first-order equation
\be
\phi' =  - \fft{h\, e^{\ft12\phi}}{a^2\, b}\, (P- Q\, e^{-\phi})\,.
\label{phifo}
\ee

   Turning to the supersymmetry variations of the gravitino, we find
that $\delta\psi_\mu=0$ in the string world-sheet directions implies
\be
\fft{c'}{h\, c} = -\fft{e^{\ft12\phi}}{4 a^2\, b}\,
  (P + Q\, e^{-\phi})\,,\label{cfo}
\ee
where $\ep$ is independent of the string world-sheet coordinates
$x^\mu$.  From the $\delta\psi_i=0$ conditions in the $S^3$ directions,
we find from $i=1,2$ and $i=3$
\bea
\fft{a'}{h\, a} &=& \fft{e^{\ft12\phi}}{4 a^2\, b}\,
(P+Q\,e^{-\phi}) - \fft{b}{2a^2}\,,\label{afo}\\
\fft{b'}{h\, b} &=& \fft{e^{\ft12\phi}}{4 a^2\, b}\, (P+Q\,
e^{-\phi}) +\fft{b}{2a^2} - \fft{(1-2k\, g)}{b} \,,\label{bfo}
\eea
respectively, where $\ep$ is independent of the Euler-angle
coordinates on $S^3$.  (We have made use of the projection conditions
(\ref{gam12proj}) and (\ref{gam1234proj}) here.)  Finally, the
variation $\delta\psi_4=0$ allows us to solve for the $r$-dependence
of the Killing spinor $\ep$.  We have
\be
\delta\psi_4 = \fft{1}{h}\, \fft{\del \epsilon}{\del r}
      + \fft{1}{8 a^2\, b}\, e^{\ft12\phi}\, (P+ Q\, e^{-\phi})\,
\Gamma_{1234}\, \ep\,,
\ee
whence, using (\ref{gam1234proj}) and comparing with (\ref{cfo}),
we find that the $r$-dependence of $\ep$ is given by
\be
\ep(r) = c^{1/2}\, \ep_0\,,
\ee
where $\ep_0$ is a constant spinor satisfying the same projection
conditions (\ref{gam12proj}) and (\ref{gam1234proj}).

{}From \eq{asol2}, \eq{bsol}, \eq{phifo} and \eq{bfo} we also find
that the constant parameters $P,g,k$ must be related as
\be 4 g\, P = k\, (1-2g\, k)\label{algcon} \ee

So far we have shown that the $F$ and $H$ field equations and
Bianchi identities, and the Killing spinor conditions are
satisfied. The remaining field equations are the Einstein and
dilaton field equations. We have verified by explicit computation
that they are also satisfied. In fact, this is not surprising
as we will show in the next section where we study the
integrability of the Killing spinor conditions.

In summary, we have a dyonic string solution given by \eq{vacans},
\eq{bsol}, \eq{asol2} and \eq{algcon}, with $h$ arbitrary and
$c$, $\phi$ determined by \eq{cfo} and \eq{phifo}. In Sec. (4.3), we
shall make a convenient choice for $h$ and study various properties
of our solution.


\subsection{Integrability of the Killing spinor equations}


In this section we shall show that once the $F$ and $H$ field
equations and Bianchi identities, and the Killing spinor
conditions are satisfied by our ansatz, the remaining Einstein and
dilaton field equations are automatically satisfied as well as a
consequence of the Killing spinor integrability conditions. As a
byproduct we will determine the full Killing spinor integrability
conditions and observe that the first order Killing spinor
equations by themselves are in general insufficient to guarantee
that all the equations of motion are satisfied.

We now determine the Killing spinor integrability conditions. For
the gravitino variation, we may take the usual commutator of
(generalized) covariant derivatives.  After considerable algebra,
we obtain
\begin{eqnarray}
\Gamma^N[\tilde D_M,\tilde D_N]&=&
-\ft12[R_{MN}-\ft14\partial_M\phi\partial_N\phi-\ft12e^{\fft12\phi}(F^2_{MN}
-\ft18g_{MN}F^2)\nonumber\\
&&\kern1cm-\ft14e^\phi(H^2_{MN}-\ft16g_{MN}H^2)
-2g^2e^{-\fft12\phi}g_{MN}]\Gamma^N\nonumber\\
&&-\ft1{48}e^{\fft12\phi}\left(\partial_{[N}H_{PQR]}-\ft34F_{[NP}F_{QR]}\right)
\Gamma^{NPQR}\Gamma_M\nonumber\\
&&-\ft1{16}e^{-\fft12\phi}\nabla^N(e^\phi H_{NPQ})\Gamma^{PQ}\Gamma_M\nonumber\\
&&-\ft18(\partial_M\phi+\ft1{12}e^{\fft12\phi}H_{NPQ}\Gamma^{NPQ}\Gamma_M)
\Delta_\phi
+\ft18e^{\fft14\phi}F_{MN}\Gamma^N\Delta_F\nonumber\\
&&-\ft1{64}\Gamma_M(e^{\fft14\phi}F_{NP}\Gamma^{NP}+8\im\,g\,e^{-\fft14\phi})
\Delta_F\,,
\end{eqnarray}
where the last two lines vanish when acting on Killing spinors.
(The quantities $\Delta_\phi$ and $\Delta_F$ are defined in
(\ref{susytrans}) and are supersymmetry transformations on $\chi$
and $\lambda$, up to unimportant numerical factors.) We see that
once the $H$ field equation, Bianchi identity and the Killing
spinor conditions are satisfied, and given that the Ricci tensor
is diagonal, the Einstein equation is then satisfied as well.

Additional integrability conditions may be derived from the
$\delta\chi$ and $\delta\lambda$ variations.  For the tensor
multiplet, we find
\begin{eqnarray}
\Gamma^M[\tilde D_M,\Delta_\phi]&=&
[\,\square\,\phi -\ft14e^{\fft12\phi}F^2 -\ft16e^\phi H^2 +8g^2e^{-\fft12\phi}]
\nonumber\\
&&-\ft16e^{\fft12\phi}\left(\partial_{[M}H_{NPQ]}-\ft34F_{[MN}F_{PQ]}\right)\Gamma^{MNPQ}
-\ft12e^{-\fft12\phi}\nabla^M(e^\phi H_{MNP})\Gamma^{NP}\nonumber\\
&&+\ft1{24}e^{\fft12\phi}H_{MNP}\Gamma^{MNP}\Delta_\phi
-\ft18\left(e^{\fft14\phi}F_{MN}\Gamma^{MN}+8\im\,g\,e^{-\fft14\phi}\right)\Delta_F\,.
\end{eqnarray}
This shows once the $H$ field equation and Bianchi identity and
the Killing spinor conditions are satisfied, then the dilaton
field equation is satisfied as well.

Finally, from the Killing spinor condition coming form the Maxwell
multiplet we find
\begin{eqnarray}
\Gamma^M[\tilde D_M,\Delta_F]&=&
e^{\fft14\phi}\partial_{[M}F_{NP]}\Gamma^{MNP}+2e^{-\fft14\phi}
[\nabla^M(e^{\fft12\phi}F_{MP})-\ft12e^\phi H_{MNP}F^{MN}]\Gamma^P\nonumber\\
&&-\ft14\Gamma^M\partial_M\phi\Delta_F+\ft12e^{\fft14\phi}F_{MN}\Gamma^{MN}
\Delta_\phi+\ft14[\Delta_\phi,\Delta_F]\,,
\end{eqnarray}
which is automatically satisfied as a result of the $F$ field
equation and the Killing spinor conditions.

The Killing spinor integrability conditions presented above can
also be used to analyze in more general situations the extent to
which they imply the field equations. We leave this to a future
work, and we next analyze the properties of our dyonic string
solution.

\subsection{The properties of the supersymmetric string solution}

In this section, we shall make a convenient choice for the
``coordinate gauge function'' $h$ and exhibit the explicit form of
the dyonic string solution, and study its salient properties such
as its behavior in various limits.  In particular, we choose $h$ so that
the solutions for $\phi$ and $c$ will be identical to those in
\cite{dulupo} for the gauge dyonic string.  This is achieved by
making the gauge choice
\be
h  =  - \fft{2 a^2\, b\, c^2}{r^3}\,,
\ee
and defining $\phi_\pm \equiv \phi \pm 4 \log c$, whereupon the
equations become diagonalised, with
\be
\phi_+' = \fft{4P}{r^3}\, e^{\ft12\phi_+}\,,\qquad
\phi_-' = -\fft{4Q}{r^3}\, e^{-\ft12\phi_-}\,.
\ee
The solutions can be written as
\be
e^{-\fft12\phi_+} = P_0 + \fft{P}{r^2}\,,\qquad
e^{\fft12\phi_-} = Q_0 + \fft{Q}{r^2}\,,
\ee
and hence
\be
e^\phi= \Big(Q_0 + \fft{Q}{r^2}\Big)\,  \Big(P_0 +
\fft{P}{r^2}\Big)^{-1}
\,,\qquad
c^{-4} =  \Big(Q_0 + \fft{Q}{r^2}\Big)\,  \Big(P_0 + \fft{P}{r^2}\Big)
\,.\label{phicsol}
\ee

    Were we indeed looking for the dyonic string solutions in the ungauged
theory, we would then solve (\ref{afo}) and (\ref{bfo}) for $a$ and
$b$, with $g=0$.  In our present case, however, we already have
the algebraic equations (\ref{bsol}) and (\ref{asol2}), which came from
solving the $F_2$ field equation and the $\delta\lambda=0$
supersymmetry condition respectively.  (Both these conditions would
have been vacuous in the dyonic string solutions in the ungauged theory.)
Thus our solution is simply given by (\ref{phicsol}), together with
the expressions for $a$ and $b$ in (\ref{asol2}) and (\ref{bsol}).
Collecting the above results, we find that the dyonic string solution of
the six-dimensional gauged $N=(1,0)$ supergravity is given by
\begin{eqnarray}
ds^2 &=&H_P^{-\fft12}H_Q^{-\fft12}\,dx^\mu\,dx_\mu
+\fft{k^2\,P}{4g^2r^6}H_P^{-\fft52}H_Q^{\fft12}\,dr^2
+\fft{k}{4g}H_P^{-\fft12}H_Q^{\fft12}\Bigl(\sigma_1^2 + \sigma_2^2
+ \fft{4g\, P}{k}\, \sigma_3^2 \Bigr)\,,\nn\\
H_\3 &=& P \, \sigma_1\wedge\sigma_2\wedge\sigma_3 - d^2x\wedge dH_Q
^{-1}\,,\qquad F_\2 = k\, \sigma_1\wedge\sigma_2\,,\nn\\
e^\phi &=&H_Q/H_P\,,
\label{eq:pqsoln}
\end{eqnarray}
where
\begin{equation}
H_Q=Q_0+\fft{Q}{r^2}\,,\qquad
H_P=P_0+\fft{P}{r^2}\,.
\end{equation}
Note that the $H_\3$ and $F_\2$ charges must satisfy the algebraic constraint
(\ref{algcon}), namely $4g\,P=k(1-2g\,k)$.

Before turning to the properties of this solution, we may examine its
relation to the dyonic string of the ungauged theory \cite{dulupo}.  To
highlight the similarities, we may reexpress the metric as
\begin{equation}
ds^2=(H_PH_Q)^{-\fft12}dx^\mu\,dx_\mu+(H_PH_Q)^{\fft12}\left[
4\xi^2\Xi^{-3}\,dr^2
+\Xi^{-1}\,r^2(\xi(\sigma_1^2+\sigma_2^2)+\sigma_3^2)\right]\,,
\label{eq:npqmet}
\end{equation}
where
\begin{equation}
\xi=\fft{k}{4gP}=(1-2g\,k)^{-1}\,,\qquad\Xi=1+\left(\fft{P_0}P\right)r^2\,.
\end{equation}
To obtain the ungauged theory, we may take the limit $g\to0$, $k\to0$ with
the $H_\3$ magnetic charge $P\to k/4g$ held fixed, so that $\xi\to1$.  The
metric, (\ref{eq:npqmet}), then approaches that of the dyonic string in
the ungauged theory, provided $\Xi\to1$.  This latter condition is perhaps
somewhat surprising, as this restriction is absent in the ungauged theory.
Its origin is apparently related to the nature of turning on both $F_\2$
and $H_\3$ flux over the squashed $S^3$.

We now return to the gauged theory, and consider the properties of the dyonic
string solution, (\ref{eq:pqsoln}).  For small $r$, the functions $H_P$ and
$H_Q$ blow up as $1/r^2$ (we take both $P$ and $Q$ positive).  Furthermore,
in this limit, we see that $\Xi\to1$.  Hence the near-horizon limit of the
string may be read off from the metric (\ref{eq:npqmet}) by taking $\Xi=1$
and retaining $\xi$ as a squashing parameter.  As a result, we see that
this limit in fact precisely yields the AdS$_3$ times squashed 3-sphere
family of solutions that we found in section 3.

Turning to the asymptotics away from the horizon, we note that some care
must be involved in handling the constant $P_0$.  For $P_0>0$, $H_P\to
\hbox{const}$ as $r\to\infty$. However $\Xi\sim r^2$ in this limit, and
this drastically modifies the asymptotics. In particular, $ds^2\sim dr^2/r^6$
at large $r$, so that the $r$ interval has a finite range.  On the other
hand, for $P_0=0$, the function $\Xi$ is identically $1$, and in this fashion
we are able to recover large distance asymptotics.  For $P_0<0$, the function
$H_P$ goes through zero when $r^2=|P/P_0|$, thus putting a natural limit on
the coordinate, $r\in(0,|P/P_0|^{1/2})$.

In fact, for either $P_0=0$ or $P_0<0$, the large distance asymptotics
originate when $H_P\to0$.  Both cases may be treated simultaneously by
changing to a new radial coordinate $\rho$, related to $r$ by
\be
P_0 + \fft{P}{r^2} = \fft{P^2}{\rho^4}\,.
\ee
We also replace the constants $Q_0$ and $Q$ by
\be
\wtd Q_0 \equiv Q_0 - \fft{Q\, P_0}{P}\,,\qquad
\wtd Q^2 \equiv Q\, P\,.
\ee
In terms of these redefined quantities, dyonic string solution of
(\ref{eq:pqsoln}) becomes
\bea
ds^2 &=& \fft{\rho^2}{P\, {\cal H}^{\ft12}}\,
dx^\mu\, dx_\mu + 16\, \xi^2 {\cal H}^{\ft12}\,
 d\rho^2 + {\cal H}^{\ft12}\, \rho^2\,
(\xi(\sigma_1^2 + \sigma_2^2)+ \sigma_3^2)\,,\nn\\
H_\3 &=& P \, \sigma_1\wedge\sigma_2\wedge\sigma_3 - d^2x\wedge d{\cal
H}^{-1}\,,\qquad F_\2 = k\, \sigma_1\wedge\sigma_2\,,\nn\\
e^\phi &=&  \fft{{\cal H}\, \rho^4}{P^2}\,,\qquad {\cal H} \equiv
\wtd Q_0 + \fft{\wtd Q^2}{\rho^4}\,.\la{48} \eea

If $\wtd Q_0$ is positive, these solutions describe everywhere
non-singular dyonic strings.\footnote{If $\wtd Q_0$ is negative,
there is a naked singularity at the value of $\rho$ for which
${\cal H}$ vanishes.  If $\wtd Q_0=0$, the metric (\ref{48})
coincides precisely with the near-horizon metric (\ref{49}), which
is nothing but AdS$_3$ times the squashed 3-sphere.}  The
previously noted horizon at $r=0$ corresponds here to $\rho=0$,
and in the near-horizon limit where $\rho \longrightarrow 0$, the
metric approaches
\be ds^2 \sim (PQ)^{\ft12}\, \Big(16\,\xi^2\fft{d\rho^2}{\rho^2} +
\fft{\rho^4}{P^2\, Q}\, dx^\mu\, dx_\mu \Big) + (PQ)^{\ft12}\,
(\xi(\sigma_1^2 + \sigma_2^2) + \sigma_3^2)\,,\la{49} \ee
while the dilaton approaches a constant; $e^\phi \longrightarrow
Q/P$.  In this limit the tensor multiplet is frozen, and the projection
condition (\ref{gam1234proj}) is lost.  As a result, the supersymmetry
at the horizon is restored to $\ft12$ of the original supersymmetry.

   At large distance, $\rho\to\infty$, the metric approaches
\be ds^2 \sim 16 \xi^2\, \wtd Q_0^{\ft12}\, \Big(d\rho^2 +
\fft1{16\xi^2}\, \rho^2\, (\xi(\sigma_1^2 + \sigma_2^2) +
\sigma_3^2)  + \fft{P\, g^2}{k^2\wtd Q_0}\, \rho^2\, dx^\mu\,
dx_\mu \Big)\,, \ee
which describes a cone over the product of a squashed $S^3$ and the
(Minkowski)$_2$ string worldsheet metric.  Unlike the gauge dyonic
string \cite{dulupo}, or for that matter typical string solitons
\cite{Duff:1994an}, the present dyonic string is not asymptotic to the
usual vacuum attributed to this gauged $N=(1,0)$ theory, namely
(Minkowski)$_4\times S^2$.  In fact, looking at the dilaton, one finds
$e^\phi\sim\rho^4(\wtd Q_0/P^2)$, which blows up asymptotically.  This
is not necessarily surprising, since the potential is of a single-exponential
form, which suggests the possibility of a domain-wall type solution with
runaway dilaton.  We note, however, that when the dilaton is active, the
$\delta\chi=0$ condition requires non-vanishing $H_\3$ for the preservation
of supersymmetry.  This indicates that domain wall solutions with
(Minkowski)$_5$ symmetry do not occur, and that the (Minkowski)$_2$ times
squashed $S^3$ geometry above is perhaps the most symmetric that may be
obtained for a domain wall configuration.

\section{Discussion}

    The construction of a new family of solutions to the gauged
$N=(1,0)$ theory suggests that there exists a wider class of
supersymmetric vacua than previously anticipated.  In particular, the
well-known (Minkowski)$_4\times S^2$ solution is but a special
singular limiting case of a larger class of solutions of the form
AdS$_3$ times squashed $S^3$.  Thus it is apparent that some form of
fine tuning is necessarily present to obtain a flat Minkowski
spacetime.  The half-supersymmetric vacua constructed in section
\ref{sec:vacsol} are parametrized by the $H_\3$ flux $P$, with
$P\longrightarrow0$ corresponding to the Minkowski limit.  For
non-zero $P$, on the other hand, the 3-form singles out three of the
six dimensions to form a 3-sphere, and one of the
(Minkowski)$_4$ dimensions is lost to become the Hopf fibre of $S^3$.

    We now turn to the implications of this result on the recent
braneworld models with football-shaped extra dimensions
\cite{Carroll:2003db,Navarro:2003vw,Aghababaie:2003wz}.  While we have
little to say about the effect of non-supersymmetric 3-branes
({\it i.e.}~deficit angles) on the cosmological constant, it should now
be evident that a fine-tuning of the bulk geometry is nevertheless required,
even before the introduction of branes into the bulk.  The vanishing of
the cosmological constant in the (Minkowski)$_4\times S^2$ background does
indeed arise after assuming an $M_4\times S^2$ symmetry of the vacuum.
However, as we have seen, supersymmetry itself is insufficient for selecting
such a symmetry, and the theory itself is perfectly content to compactify
spontaneously with an AdS$_3$ times squashed $S^3$ geometry.  This feature is
similar to what occurs in eleven-dimensional supergravity, where
(Minkowski)$_{11}$ may be obtained as a limiting case of the
AdS$_4\times S^7$ Freund-Rubin compactification \cite{Freund:1980xh}.

    In fact, obtaining a naturally small cosmological constant in these
braneworld models comes to the issue of balancing the $F_\2$ and $H_\3$
fluxes against the six-dimensional potential.  Reduced onto the braneworld,
this mechanism essentially replaces the effective cosmological constant
by a dynamical variable.  This is similar to earlier ideas where the
cosmological constant is replaced by a four-form field strength in four
dimensions \cite{Hawking:hk,Duff:1989ah}.  In such models, the four-form
itself has no local dynamics, but may take on appropriate values so as to
cancel the background vacuum energy.  In the end, however, this effective
four-form may take on a range of values (quantized in the case of M-theory
\cite{Bousso:2000xa}), and one is again reduced to an anthropic argument
to explain the smallness of the cosmological constant.

    More generally, we would like to investigate whether any
additional supersymmetric vacua of this gauged $N=(1,0)$ theory may
exist.  In addition to the AdS$_3$ times squashed $S^3$ backgrounds
preserving $\fft12$ of the original supersymmetry, we have also identified
a class of dyonic string solutions preserving $\fft14$ of the original
supersymmetry.  Although we believe we have essentially exhausted the
possibility of static vacua, a more systematic treatment would be
necessary.  Since the field content of the theory as well as its
multiplet structure is relatively simple, it may be amenable to an
analysis similar to that which was performed in
\cite{Gauntlett:2002nw,Gauntlett:2003fk} for the case of minimal
(gauged and ungauged) supergravity in five dimensions.  Some
preliminary analysis is currently under way.

    Finally, it remains an unresolved issue whether the minimal gauged
supergravity may be obtained from a higher-dimensional theory.  While
ungauged $N=(1,0)$ theories are easily obtained from, for example, heterotic
strings reduced on $K3$, obtaining a chiral gauged supergravity from higher
dimensions appears to be a more difficult task.  One may naturally obtain
gauged supergravities from compactifications with fluxes, and in fact
gauged $N=2$ supergravity in five dimensions may be obtained by a flux
compactification on $K3\times S^1$ \cite{Louis:2001uy}.  However it is
not clear that this could be lifted up to six dimensions.  In fact, it
is argued in \cite{Louis:2001uy} that no background fluxes can be turned on
for the heterotic on $K3$ reduction to six dimensions in itself.

    Alternatively, the (Minkowski)$_4\times S^2$ background with monopole
configuration is suggestive of a seven-dimensional interpretation with
(Minkowski)$_4\times S^3$ vacuum, where the $S^3$ may be viewed as a $U(1)$
bundle over $S^2$ \cite{salsez}.  However attempts at reducing gauged $N=2$
supergravity in seven dimensions to yield the gauged $N=(1,0)$ theory in six
has so far proven unsuccessful.  Note that the ungauged $N=(1,0)$ theory may
be obtained in this fashion through a braneworld reduction for
both pure supergravity \cite{Lu:2000xc} and supergravity coupled to a single
tensor multiplet \cite{Liu:2002vi}.  However, to reduce to the gauged theory,
one needs to retain at least a vector multiplet in the braneworld reduction.
So far this has only been accomplished in the bosonic sector
\cite{Liu:2002vi}.

In particular, the gauged theory involves a chiral six-dimensional
gravitino charged under the Maxwell field.  Ordinarily, this Maxwell
field would be labeled as a graviphoton, namely a superpartner of the
graviton and gravitino.  However, in the $N=(1,0)$ theory, it is actually
part of an ordinary Maxwell multiplet and has a single spin-$\fft12$
superpartner.  For a braneworld reduction from seven dimensions, this
Maxwell field can only arise as the $U(1)$ component of the $SU(2)$
graviphotons, as one needs to retain the gauged $R$-symmetry of the
gravitino (and there is no $U(1)$ isometry in the reduction of the metric).
However this leads to an apparent incomplete gravitino multiplet in
the reduction, and not the requisite Maxwell multiplet \cite{Liu:2002vi}.

More importantly, it is not clear how one obtains a chiral charged
gravitino from dimensional reduction of ten or eleven dimensional
supergravities.  Since the higher-dimensional theory necessarily involves
uncharged gravitini, a gauged $R$-symmetry must somehow arise from the
dimensional reduction.  In this case, one would have to ensure chirality,
either through singularities or non-perturbative effects, or perhaps by
an overlooked mechanism in the braneworld reduction.  An alternative
approach would be to obtain chirality by consistent truncation of a
non-chiral theory, such as the $N=(1,1)$ gauged supergravity in six
dimensions.  However, a straightforward truncation of the bosonic sector
of the $N=(1,1)$ supergravity yields an incorrect scalar potential.
Furthermore, it is not clear that the fermion sector (and especially the
resulting $N=(1,0)$ gravitino multiplet) may be consistently truncated.

Thus the dyonic string solutions that we have obtained are not obviously
related to any of the
well-known objects in M-theory, although they do share similarities with
the dyonic strings of the ungauged theory \cite{dulupo}.  Of course, the
minimal theory considered here \cite{salsez} is anomalous, and must be
supplemented with additional matter for consistency
\cite{Randjbar-Daemi:wc,Salam:1985mi,Bergshoeff:1986hv}.
It is presumably the anomaly-free models, if any, that may be lifted to
higher dimensions.


\section*{Acknowledgments}

J.T.L. wishes to thank D.~Chung and M.~Porrati for informative discussions,
and acknowledges the hospitality of the theory group at the Rockefeller
University, where part of this work was done.  R.G., C.N.P. and
E.S. are grateful to the Michigan Center for Theoretical Physics, and
C.N.P. is grateful to the Cambridge Relativity and Gravitation Group,
for hospitality during the course of this work.


\end{document}